\documentstyle[12pt]{article}
\newcommand{\be}{\begin{equation}}
\newcommand{\ee}{\end{equation}}
\newcommand{\ba}{\begin{eqnarray}}
\newcommand{\ea}{\end{eqnarray}}
\newcommand{\baa}{\begin{eqnarray*}}
\newcommand{\eaa}{\end{eqnarray*}}
\newcommand{\bb}{\begin{thebibliography}}
\newcommand{\eb}{\end{thebibliography}}

\newcommand{\bi}[1]{\bibitem{#1}}
\begin {document}

\rightline{RUB-TPII-22/98; TPR-98-38}

\begin{center}
{\Large \bf The intrinsic charm contribution to the proton spin}\\[1cm]
M.V. Polyakov$^{1,2}$, A. Sch\"afer$^3$, O.V. Teryaev$^4$
\\[0.3cm]
{$^1$\it Petersburg Nuclear Physics Institute,
188350 Gatchina, Russia. }\\[0.3cm]
{$^2$\it Institut f\"ur Theoretische Physik II,\\
Universit\"at Bochum, D--44780 Bochum }
\\[0.3cm]
{$^3$\it Institut f\"ur Theoretische Physik\\
Universit\"at Regensburg\\
D-93040 Regensburg, Germany }\\[0.3cm]
{$^4$\it Bogoliubov Laboratory of Theoretical Physics\\
Joint Institute for Nuclear Research, Dubna\\
141980 Russia}\\[0.3cm]

\end{center}
\begin{abstract}
The charm quark contribution to the first moment of
$g_1(x,Q^2)$ is
calculated using a heavy mass expansion of the divergence
of the singlet
axial current. It is shown to be small.
\end{abstract}

\newpage
The size of a possible intrinsic charm contribution in the proton
has been the topic of intensive discussions \cite{C1,C2,C3,C4}
for many years. It is
therefore a natural question to investigate the {\em polarized}
intrinsic charm distribution in the nucleon \cite{C5}.
Recently one of us and collaborators argued \cite{Maxim1}
that
earlier treatments of the polarized charm contribution
to the $\eta$ and $\eta'$
\cite{HZ,AM} were incorrect. In this contribution we extent
and adopt that analysis to the nucleon. More precisely we
shall focus on the intrinsic charm contribution to the
first moment of the spin structure function $g_1(x,Q^2)$.
This is known to be intimately related
to the gluonic axial anomaly \cite{ET,AR,CCM,MS}.
It may be expressed as forward limit of $G_{A}^{(0)}(t)$,
the form
factor in the proton matrix element
of the singlet axial current
\begin{eqnarray}
&&\!\!\!\!\!\!\!\!\!\!\!\!\!\!\!\!\langle N(p_2,{\lambda}_2)| j_{5\mu}^{(0)}(0)
|N(p_1,{\lambda}_1) \rangle \nonumber\\
&&\qquad \qquad =\bar u_{N}^{({\lambda}_2)}(p_2) \left(G_{A}^{(0)}(t){\gamma}_
{\mu}{\gamma}_5-G_{P}^{(0)}(t)q_{\mu}{\gamma}_5 \right) u_{N}^{({\lambda}_1)}(
p_1), \label{formfac}
\end{eqnarray}
where $q=p_2-p_1$ and $t=q^2$.
The singlet pseudoscalar form factor does
not acquire a Goldstone pole at $t=0$, even in the chiral limit,
contrary to the matrix elements of the octet currents.
In this limit, there exist eight massless pseudoscalar
mesons serving as Goldstone bosons. However, the ninth pseudoscalar,
the ${\eta}'$-meson, remains massive, due to the mixing with the QCD
ghost pole.

This fact allows to relate the forward matrix element
of the axial current to the (slightly) off-forward one of
its divergence:

\begin{eqnarray}
&&\!\!\!\!\!\!\!\!\!\!\!\!\!\!\!\!
\lim_{t\to 0}\langle N(p_2,{\lambda}_2)|{\partial}^{\mu}
 j_{5\mu}^{(0)}(0)
|N(p_1,{\lambda}_1) \rangle \nonumber\\
&&\qquad \qquad = 2 m_N G_{A}(0)
\bar u_{N}^{({\lambda}_2)}(p_2)
{\gamma}_5 u_{N}^{({\lambda}_1)}(p_1), \label{formfacdiv}
\end{eqnarray}
$m_N$ being the proton mass.
The divergence of the singlet axial current in turn
contains a normal and an anomalous piece,
\begin{equation}
{\partial}^{\mu}j_{5\mu}^{(0)}=2i\sum_{q}m_{q}\bar q
{\gamma}_5q-\left(\!  \frac{ N_{f}{\alpha}_{s}}{4\pi}\! \right)G_{\mu
\nu}^a\widetilde G^{\mu \nu,a}, \label{div} \end{equation}
where
$N_{f}$ is the number of flavours.  The two terms at the r.h.s. of the
last equation are known to cancel in the limit of infinite quark mass
\cite{CCM,MS,EST}.  This is the so-called cancellation of physical and
regulator fermions, related to the fact, that the anomaly may be
regarded as a usual mass term in the infinite mass limit, up to a sign,
resulting from the subtraction in the definition of the regularized
operators.

Consequently, one should expect, that the contribution of
infinitely
heavy quarks
to the first moment of $g_1$ is zero.
This is exactly what happens in a perturbative calculation of the
triangle anomaly graph \cite{MS}.
One may wonder, what is the size of this
correction for large, but finite masses
and how does it compare with the purely perturbative
result. To answer this
question, one should calculate the r.h.s. of (\ref{div}) for heavy fermions.
The leading coefficient is of the order $m^{-2}$,
and its calculation  was addressed  recently
by two groups \cite{HZ} and \cite{AM} who
came up with results
differing by a factor of six. However, the operator
$f_{abc}G_{\mu \nu }^a\tilde G_{\nu \alpha }^bG_{\alpha \mu
}^c$ appearing in both treatments, does not satisfy some basic
properties, such that both calculations seem to be flawed.

i) It is not a divergence of a local operator,
therefore it is not clear that its forward matrix element
(\ref{formfacdiv}) will vanish.\\
iii) It makes no contact with the calculation of the triangle diagram
in momentum space \cite{Adler,MS} being
essentially non-abelian.\\

The recent contribution \cite{Maxim1} corrected this result and
arrived at the expression:

\begin{equation}
{\partial}^{\mu}j_{5\mu}^{c}
 =\frac{\alpha_s}{48\pi m_c^2}\partial^\mu R_\mu
\,
\label{theres}
\end{equation}
where
\begin{equation}
R_\mu =\partial_\mu \left( G_{\rho \nu }^a\tilde
G^{\rho \nu, a }\right)
-4\left( D _\alpha
G^{\nu \alpha }\right)^a \tilde G_{\mu \nu }^a
\; .  \label{res-2}
\end{equation}
[Here we use the conventions: $\gamma_5=i\gamma^0\gamma^1\gamma^2\gamma^3$
and $\varepsilon_{0123}=1$]
This result is an explicit 4-divergence and
has the Abelian limit.
Moreover, this result can also be obtained by a
$1/m$ expansion of the triangle diagram contribution.
Actually the result of \cite{Maxim1} demonstrates that
in order $m_c^{-2}$ the entire result (\ref{res-2})
can be restored from the venerable triangle diagram. The
diagrams with larger number of ``legs''
give only contributions to the non-abelian part of the
result (\ref{res-2}). Indeed, computing the forward matrix element
of operator (\ref{res-2})
between two virtual gluon states we get the following expressions:

\begin{equation}
\langle p|\frac{\alpha_s }{48\pi m_c^2}R_\mu|p\rangle=
-i\frac{\alpha_s }{12\pi } \varepsilon_{\mu\nu\lambda\rho}
e^\nu  e^{*\rho}p^\lambda \frac{p^2}{m_c^2}\, .
\label{treugolnik}
\end{equation}
On other hands the result of a calculation of the triangle diagram
with massive fermions
(see e.g. \cite{CCM}) has a form:
\begin{eqnarray}
\nonumber
\langle p|\bar c\gamma_\mu \gamma_5 c|p\rangle&=&
i\frac{\alpha_s }{2\pi } \varepsilon_{\mu\nu\lambda\rho}
e^\nu  e^{*\rho}p^\lambda
\Bigl\{
1-\int_0^1 dx \frac{2 m_c^2(1-x)}{m_c^2-p^2x(1-x)}
\Bigl\} \\
&=&
-i\frac{\alpha_s }{12\pi } \varepsilon_{\mu\nu\lambda\rho}
e^\nu  e^{*\rho}p^\lambda\frac{p^2}{m_c^2} + O(\frac{1}{m_c^4})\, .
\end{eqnarray}
This expression coincides exactly with the result (\ref{treugolnik}).
In order to complete the proof it is enough to consider
the off-forward matrix element of the operator (\ref{theres})  between
two gluons at zero virtuality and compare the result with the expression
for the triangle diagram for
$\langle p'|\partial^\mu \bar c
\gamma_\mu \gamma_5 c|p\rangle$. It is easy to check that again the results
coincide.

The proton matrix element of $R_\mu$ takes a form analogous to
that of (\ref{formfac}) \begin{eqnarray}
&&\!\!\!\!\!\!\!\!\!\!\!\!\!\!\!\!\langle N(p_2,{\lambda}_2)| R_{\mu} (0)
|N(p_1,{\lambda}_1) \rangle \nonumber\\
&&\qquad \qquad =\bar u_{N}^{({\lambda}_2)}(p_2) \left(G_{A}^{R}(t){\gamma}_
{\mu}{\gamma}_5-G_{P}^{R}(t)q_{\mu}{\gamma}_5 \right) u_{N}^{({\lambda}_1)}(
p_1), \label{formfacR}
\end{eqnarray}
It is crucial, that because of the explicit gauge invariance of $R_\mu$
the zero mass ghost pole does not contribute.
This make an apparent difference with respect to the massless case,
when the divergence of the gauge-dependent topological current $K_\mu$ appears
and the ghost pole contribution does not allow to deduce the relation between
the matrix elements of the currents starting from the relation for their
divergencies \cite{EST}.
In the case under investigation only the contribution of
the massive $\eta^{'}$ meson may appear so that
\begin{eqnarray}
&&\!\!\!\!\!\!\!\!\!\!\!\!\!\!\!\!
\lim_{t \to 0}
\langle N(p_2,{\lambda}_2)|{\partial}_{\mu}
 R_{\mu}(0)
|N(p_1,{\lambda}_1) \rangle \nonumber\\
&&\qquad \qquad = 2m_N G_{A}^R (0)
\bar u_{N}^{({\lambda}_2)}(p_2)
{\gamma}_5 u_{N}^{({\lambda}_1)}(p_1), \label{formfacdivR}
\end{eqnarray}

The contribution of charm to the forward matrix element can be obtained
by substituting
(\ref{formfac}, \ref{formfacR}) into the proton matrix elements
of (\ref{theres}), giving in the forward limit.
\begin{equation}
\langle N(p,{\lambda})| j_{5\mu}^{(c)}(0)
|N(p,{\lambda}) \rangle  =
{\alpha_s\over 48 \pi m_c^2}
\langle N(p,{\lambda})| R_{\mu}(0)
|N(p,{\lambda}) \rangle
 \label{jc}
\end{equation}

In deriving this expression we used
(\ref{formfacdiv}, \ref{formfacdivR}).
Note that the first term in $R_{\mu}$ does
not contribute to the forward
matrix element because of its gradient form,
while the contribution of the
second one is rewritten, by making use of the equation
of motion, as
matrix element of the operator
\begin{eqnarray}
\langle N(p,{\lambda})| j_{5\mu}^{(c)}(0)
|N(p,{\lambda}) \rangle&=\frac{\alpha_s}{12 \pi m_c^2}
\langle N(p,{\lambda})|
g\, \sum_{\scriptstyle{\rm f=u,d,s}}
\bar \psi_f\gamma_\nu \tilde G_{\mu}^{~~\nu} \psi_f | N(p,{\lambda}) \rangle \
\nonumber \\
&\equiv
\frac{\alpha_s}{12 \pi m_c^2} 2 m_N^3 s_{\mu} f^{(2)}_S,
\label{sfc}
\end{eqnarray}
The parameter $f^{(2)}_S$ was determined before in calculations
of the power corrections to the first moment
of the singlet part of $g_1$ part of which is given by
exactly the quark-gluon-quark matrix element we got.
Note that within our $1/m_c$ approximation the $c$ contribution
to the flavour sum can be neglected.
QCD-sum rule calculations gave
$f^{(2)}_S=\frac{9}{5}(
f^{(2)}({\rm proton})+f^{(2)}({\rm neutron}))=0.09$\footnote{
Note that here we use a convention for the $\varepsilon$-tensor
which differs by sign from that of \cite{ES}}
\cite{ES},
estimates using the renormalon
approach led to
$f^{(2)}_S=\pm 0.02 $
\cite{ES2} and calculations in the instanton model of the QCD vacuum
give a result very close to that of QCD sum rule \cite{ES}
\cite{Maxim2}.

Inserting these numbers we get finally for  the charm axial
constant the estimate
\begin{equation}
\bar G_A^c(0)=
-\frac{\alpha_s}{12\pi} f^{(2)}_S (\frac{m_N}{m_c})^2\approx
-5 \cdot 10^{-4}
\label{gc}
\end{equation}
with probably a 100 percent  uncertainty
(see e.g. \cite{Ioffe}).
As the mass term in the triangle diagram is coming from the region
of transverse
momenta of the order $m_c$, this should be the correct scale of both
$\alpha_s$ and $f^{(2)}_S$. Because this scale is not
far from the typical
hadronic scale at which $f^{(2)}_S$ was estimated we can neglect
evolution effects.

Note that this contribution is of non-perturbative origin
(therefore we call it intrinsic),
so that
it is sensitive to large distances, as soon
as the factorization scale is larger than $m_c$. If the scale
is also larger than
$m_b$, one can immediately conclude that
the non-perturbative bottom
contributions is further  suppressed by the
factor $(m_c/m_b)^2 \sim 0.1$.

Let us note, that the naive application of our
approach to the case of strange
quarks gives for their contribution to the first moment of $g_1$
roughly $-5 \cdot 10^{-2}$, which is compatible with the experimental data.
The possible applicability of a heavy quark expansions for strange quarks
in a similar problem
was discussed earlier
\cite{BG} in the case of the vacuum condensates of heavy quarks.
That analyses was also related to
the anomaly equation for heavy quarks,
however,  for the trace anomaly, rather than the axial one.\\

Let us summarize: We have related the
non-perturbative contribution of charm quarks to the
nucleon spin (at scale $m_c$) to the singlet twist-4
coefficent appearing e.g. in the Ellis-Jaffe sum rule. Numerically
it is found to be very small, contrary to the suggestion of
\cite{C5,AM}.
We would like to note that in a recent paper \cite{BvN}
it was shown that also the
perturbative $\Delta c$ contribution is very small. We see this as further
support for our result.

\vskip 1 cm
Acknowledgement:\\
We thank K. Goeke, M. Franz, M. Maul, S.V. Mikhailov, P.V. Pobylitsa and 
E. Stein for helpful discussions.
O.V.T. was supported by the Russian
Foundation for the Basic Research (grant 96-02-17631),
the Graduiertenkolleg Erlangen-Regensburg (DFG) and Bochum University.
M.V.P. is grateful to Klaus Goeke for encouragement and
kind hospitality at Bochum University.
The work of M.V.P.
has been supported in parts by a joint grant of the Russian
Foundation for Basic Research (RFBR) and the Deutsche
Forschungsgemeinschaft (DFG) 436 RUS 113/181/0 (R), by the BMBF grant
RUS-658-97 and the COSY (J\"ulich). A.S. acknowledges support from BMBF.

\bb{99}
\bi{C1} S.J. Brodsky, P. Hoyer, C. Peterson and N. Sakai,
Phys. Lett. B93 (1980) 451\\
S.J. Brodsky and C. Peterson, Phys. Rev. D23 (1981) 2745
\bi{C2} E. Hoffmann and R. Moore, Z.Phys. C20 (1983) 71
\bi{C3} B.W. Harris, J. Smith and R. Vogt, Nucl. Phys. B461 (1996) 181
\bi{C4} G. Ingelman and M. Thunman, Z. Phys. C73 (1997) 505
\bi{C5} I. Halperin and A. Zhitnitsky, `Polarized intrinsic
charm as a possible solution to the proton spin problem',
hep-ph/9706251; A.Blotz and E.~Shuryak
`Instanton-induced charm contribution to polarized deep-inelastic
scattering', hep-ph/9710544.
\bi{Maxim1} M. Franz, P.V. Pobylitsa, M.V. Polyakov and K. Goeke,
`On the heavy quark mass expansion for the operator
$\bar Q \gamma_5 Q$ and the charm content of $\eta$, $\eta'$,
hep-ph/9810343 preprint
\bi{HZ} I. Halperin and A. Zhitnitsky, Phys. Rev. D56 (1997) 7247
\bi{AM} F. Araki, M. Musakhanov and H. Toki, `Axial currents
of virtual charm in light quark processes',
hep-ph/9808290 preprint
\bi{ET} A.V. Efremov and O.V. Teryaev, Report JINR-E2-88-287,
Czech.Hadron Symp.1988, p.302.
\bi{AR} G. Altarelli and G.G. Ross, Phys. Lett. B212 (1988)
391.
\bi{CCM} R.D. Carlitz, J.C. Collins and A.H. Mueller, Phys. Lett. B214
(1988) 229
\bi{MS} L. Mankiewicz and A. Sch\"afer,
Phys. Lett. B242 (1990) 455
\bi{EST} A.V. Efremov, J. Soffer and
O.V. Teryaev, Nucl.Phys. B346 (1990) 97
\bi{Adler} S.L. Adler, Phys.Rev. 177 (1969) 2426
\bi{ES} E. Stein, P. Gornicki, L. Mankiewicz and A. Sch\"afer,
Phys. Lett. B353 (1995) 107
\bi{ES2} E. Stein, M. Maul, L. Mankiewicz, and A. Sch\"afer,
'Renormalon model predictions for power-corrections to flavour singlet
deep inelastic structure functions', hep-ph/9803342 preprint
\bi{Maxim2} J. Balla, M.V. Polyakov and C. Weiss, Nucl. Phys. B510
(1998) 327
\bi{Ioffe} B.L. Ioffe,
Phys. Atom. Nucl. {\bf 60} (1997) 1707.
\bi{BG}
M.A. Shifman, A.I. Vainshtein and V.I. Zakharov,
Nucl.Phys. B147 (1979) 385 (section 6.8);
D.J. Broadhurst and S.C. Generalis, Phys. Lett. B139 (1984) 85.
\bi{BvN} J. Bluemlein and W. L. van Neerven,
`Heavy flavor contribution to the deep inelastic scatterint sum rules',
hep-ph/9811351
\eb

\end{document}